# MODE I AND MODE II STRESS INTENSITY FACTORS AND DISLOCATION DENSITY BEHAVIOUR IN STRAIN GRADIENT PLASTICITY

V. Shlyannikov[1*], E. Martínez-Pañeda[2], A. Tumanov[1] and R. Khamidullin[1]

[1]*Institute of Power Engineering and Advanced Technologies*
*FRC Kazan Scientific Center, Russian Academy of Sciences, Kazan, Russian Federation*
[2]*Department of Civil and Environmental Engineering, Imperial College London, London SW7 2AZ, UK*

**Abstract.** In this study, we use the mechanism-based strain gradient plasticity theory to evaluate both crack tip dislocation density behaviour and the coupled effect of the material plastic properties and the intrinsic material length on non-linear amplitude factors. The two planar classical stress-strain states are examined, namely, plane strain and plane stress, both under pure mode I and pure mode II loading conditions. The constitutive relations are based on Taylor's dislocation model, which enables gaining insights into the role of the increased dislocation density associated with large gradients in plastic strain near cracks. The material model is implemented in a commercial finite element (FE) software package using a user subroutine, and the nonlinear stress intensity factors (SIF) are evaluated as a function of the intrinsic material length, characterising the scale at which gradient effects become significant. As a result of the FE calculations of dislocation density distributions, the effects of both the fracture mode and the stress-strain state are determined. In pure mode I, the geometrically necessary dislocation (GND) density is located symmetrically with respect to the blunted crack tip. On the contrary, under pure mode II, the GND density becomes concentrated in the blunted and sharp parts of the crack tip. In this case, fracture initiation is shown to be likely to occur near the blunted region of the crack tip, where both the stress triaxiality and the GND density are at their maximum. The relation between the equilibrium state of dislocation densities and the intrinsic material length as well as the plastic SIF as a function of the work hardening exponent is discussed.

## 1. Introduction

Strain gradient plasticity (SGP) models have received significant attention in the past three decades [1,2]. By incorporating the role of plastic strain gradients, and their associated length scale parameters, SGP models have enabled capturing the size effects observed in metals at small scales, as well as regularising otherwise ill-posed boundary value problems at the onset of material softening. Several strain gradient models have been developed to numerically capture scale size effects [3-8]. Based on dislocation analysis, Fleck and Hutchinson [3,4] developed a strain gradient theory, where one or several length parameter(s), $l$, were introduced to balance the dimensions of strains and strain gradients. This length was considered an intrinsic material length, dependent on the microstructure of the material. Fleck and Hutchinson's [3, 4] theory assumes that material hardening is influenced by both GNDs and statistically stored dislocations (SSDs). This dislocation hardening interpretation was first proposed by Ashby [9], who stated that the resistance to dislocation motion is caused by secondary dislocations piercing the slip planes, which multiply during plastic deformation and increase slip resistance. In that study, a distinction was made between SSDs, accumulating during uniform deformation, and GNDs, which are required to preserve lattice compatibility. Using tension and torsion on polycrystalline copper wires, Liu et al. [10] compared the experimental and theoretical evaluations of three phenomenological SGP theories to demonstrate that the size effects observed in plastic flow were primarily due to the GND density generated as a result of plastic strain gradients. In continuum-like models, one way of establishing a theoretical connection with the underlying dislocation density is by means of Nye's dislocation tensor [11,12]. Fleck et al. [13] assumed that Nye's dislocation density tensor provided a direct measure of the number of GNDs associated with slip gradients. Evers et al. [14] indicated that owing to heterogeneity, the GND density can be computed to satisfy the requirement of restoring the compatibility of the crystallographic lattice.

Calcagnotto et al. [15] investigated two methodologies to calculate the GND density from electron backscatter diffraction (EBSD) data. Based on the assumption of a series of twist subgrain boundaries in the cylinder, each containing two perpendicular arrays of screw dislocations, the misorientation angle is considered to be related to the GND density. The second method to evaluate GND densities is based on the calculation of the full dislocation density tensor. GNDs are characterised by the Burgers vector $b$ (slip direction) and the tangent vector $t$ (dislocation line direction). An experimental study of the dislocation density behaviour under low-cycle deformation was conducted by Jiang et al. [16]. EBSD was combined with high spatial resolution digital image correlation (HR-DIC) to explore full-field plastic strain distributions, together with finite element (FE) modelling, to understand the microcrack nucleation mechanisms. It is found that the geometry of the inclusion plays a significant role in increasing local stress; however, the nucleation of cracks is not exclusively dependent on the inclusion geometry, and a major role is also played by local GND accumulation and plastic slip. Recently, Das et al. [17] provided a comprehensive review of the theoretical background related to the computation of Nye's dislocation tensor and the underlying GND density in plastically deformed materials.

The role of GNDs in crack tip mechanics has received significant theoretical attention [18-20]. The plastic zone adjacent to the crack tip is physically small and exhibits steep plastic strain gradients, which must be accommodated by an increased GND density that provides an additional source of work hardening and strengthening. Relative to continuum models considering only the role of SSDs, such as von Mises plasticity theory, these dislocation hardening and strengthening mechanisms lead to a significant increase in the crack tip stress level. This stress elevation provides a mechanistic rationale for brittle fracture in the presence of plasticity [21]. In the context of stationary crack tip analyses, Qu et al. [18] calculated the contours of both GND and SSD densities at the crack tip. The density of GNDs was found to be large around the crack tip but rapidly decreased away from it, but the density of SSDs decreased more gradually at the same distance. Using both mechanism-based and phenomenological SGP theories, Martínez-Pañeda and Niordson [22] demonstrated localized strain hardening near crack tips, promoted by GNDs. Recently, Martínez-Pañeda and Fleck [23] showed that higher-order SGP models predicted an elastic stress state very close to the crack tip, akin to a dislocation-free zone. In terms of crack growth resistance, Wei and Hutchinson [24] and Martínez-Pañeda et al. [25] used strain gradient plasticity to show that conventional plasticity analyses overestimated the crack growth resistance of metals.

Only a few studies related to both mode I and mode II based on a strain-gradient-dependent plasticity model have been reported in the literature. Xia and Hutchinson [26] presented plain strain results for the pure mode I and mode II fields for a limited class of solids whose dependence on strain gradients was through a single invariant curvature. Huang et al. [27] applied the SGP theory to investigate the asymptotic field (stresses and coupled stresses) near a mixed-mode crack tip in elastic and elastic-plastic materials with strain gradient effects. More recently, Goutianos [28] considered a finite strain version of Fleck and Hutchinson [29] theory to identify the relationship between the material length scales and the plastic deformation at the crack tip of Mode I and mixed-mode cracks.

The aforementioned investigations reveal the significant influence of strain gradients on a wide range of fracture mechanics problems. However, there is a need to systematically characterise the role of plastic gradients in general mixed-mode fracture, and to establish a connection with the underlying dislocation hardening mechanisms. To achieve this, we use the conventional mechanism-based strain gradient (CMSG) plasticity theory, which provides a relationship between macroscopic quantities such as strains and strain gradients with mesoscopic variables such as SSD and GND densities, *via* Taylor's dislocation model. Differences between two classical stress-strain states, namely plane strain and plane stress, are examined under pure mode I and pure mode II loading conditions. Nonlinear amplitude factor solutions are determined across a wide range of material work hardening conditions, and the role of the intrinsic material length and a coupled effect of these parameters are identified and discussed.

## 2. Governing equations of conventional mechanism-based strain gradient plasticity theory

Huang et al. [30] formulated the conventional theory of mechanism-based SGP, which is a lower-order theory based on Taylor's [31] dislocation model. In this theory, the strain gradient effect is induced via the incremental plastic modulus and thus the model is free from the requirements of higher-order approaches, facilitating numerical implementation. Herein, only a brief description of the theory is provided. For more details, the reader is referred to the work by Huang et al. [30]. Both CMSG plasticity and its higher-order counterpart, usually referred to as MSG plasticity, are based on Taylor's dislocation model. Accordingly, the shear flow stress $\tau$ is defined as a function of dislocation density $\rho$, Burgers vector $b$, and the shear modulus $\mu$ as follows:

$$\tau = \alpha \mu b \sqrt{\rho}, \tag{1}$$

where $\alpha$ is an empirical coefficient ranging from 0.3 to 0.5. A value of 0.5 is assumed throughout this work. The dislocation density $\rho$ is composed of the density $\rho_S$ for SSDs and the density $\rho_G$ for GNDs, such that

$$\rho = \rho_S + \rho_G \tag{2}$$

The SSD density is related to the flow stress and the material stress-strain curve in uniaxial tension, where gradient effects are non-existent, such that

$$\rho_S = \left[\sigma_{ref} f\left(\varepsilon^P\right) / M \alpha \mu b\right]^2 \tag{3}$$

The GND density is related to the effective plastic strain gradient $\eta^P$, by

$$\rho_G = \bar{r} \frac{\eta^P}{b} \tag{4}$$

where $\bar{r}$ is the Nye factor, which is approximately 1.90 for face-centered-cubic polycrystals. The measure of the effective plastic strain gradient $\eta^P$ was introduced by Gao et al. [6] in the form of three quadratic invariants of the plastic strain gradient tensor $\eta_{ijk}^p$ as:

$$\eta^P = \left( c_1 \eta_{iik}^p \eta_{jjk}^p + c_2 \eta_{ijk}^p \eta_{ijk}^p + c_3 \eta_{ijk}^p \eta_{kij}^p \right)^{1/2}. \tag{5}$$

The coefficients were determined to be $c_1 = 0$, $c_2 = 1/4$, and $c_3 = 0$ from three dislocation models for bending, torsion, and void growth, respectively. The components of the plastic strain gradient tensor $\eta_{ijk}^p$ are computed from the plastic strain tensor $\varepsilon_{ij}^p$ as:

$$\eta_{ijk}^p = \varepsilon_{ik,j}^p + \varepsilon_{jk,i}^p - \varepsilon_{ij,k}^p \tag{6}$$

The tensile flow stress is related to the shear stress through Taylor's factor $M$ such that, considering Eqs. (1)-(3),

$$\sigma_{flow} = M\tau = M\alpha\mu b \sqrt{\rho_S + \bar{r}\frac{\eta^P}{b}}, \tag{7}$$

where $M$ is taken to be equal to 3.06 for fcc metals. Considering the definition for the SSD density in Eq. (3), one can reformulate the flow stress as

$$\sigma_{flow} = \sigma_{ref}\sqrt{f^2(\varepsilon^P) + l\eta^P}, \tag{8}$$

where

$$l = 18\alpha^2 \left(\mu/\sigma_{ref}\right)^2 b \tag{9}$$

is the intrinsic material length in CMSG plasticity For metallic materials, the internal material length typically ranges between 1 and 10 µm [32]. Assuming a power-law hardening relation, a potential choice for the reference stress $\sigma_{ref}$ and the material function $f(\varepsilon^p)$ is given by

$$\sigma = \sigma_{ref} f(\varepsilon^p) = \sigma_y \left(\frac{E}{\sigma_y}\right)^N \left(\varepsilon^p + \frac{\sigma_y}{E}\right)^N \tag{10}$$

Here, $\sigma_y$ denotes the initial yield stress, and $N$ is the plastic work hardening exponent ($0 \leq N < 1$).

The constitutive prescriptions provided so far are common to both MSG (higher-order) and CMSG (lower-order) plasticity models. In the latter, the use of higher-order stresses is avoided by using a viscoplastic approach. Thus, the plastic strain rate is defined as a function of the effective stress $\sigma_e$, rather than its rate, as follows:

$$\dot{\varepsilon}^{p} = \dot{\varepsilon}\left(\frac{\sigma_{e}}{\sigma_{flow}}\right)^{m} = \dot{\varepsilon}\left[\frac{\sigma_{e}}{\sigma_{ref}\sqrt{f^{2}(\varepsilon^{p})+l\eta^{p}}}\right]^{m} \qquad (11)$$

where $m$ is the rate-sensitivity exponent. Values of $m$ larger than 5 have been shown to provide a response significantly close to the rate-independent limit. In this study, $m = 20$ was selected to simulate a rate-independent behaviour. Accordingly, the governing equations are therefore essentially the same as those in conventional plasticity and the plastic strain gradient comes into play through the incremental plastic modulus. Thus, the material Jacobian can be readily obtained from the following expression,

$$\dot{\sigma}_{ij} = K\dot{\varepsilon}_{kk}\delta_{ij} + 2\mu\left[\dot{\varepsilon}'_{ij} - \frac{3\dot{\varepsilon}}{2\sigma_{e}}\left(\frac{\sigma_{e}}{\sigma_{flow}}\right)^{m}\sigma'_{ij}\right]. \qquad (12)$$

where $\dot{\varepsilon}'_{ij}$ is the deviatoric strain rate tensor. As with other continuum strain gradient plasticity models, the CMSG theory is intended to model a collective behaviour of dislocations and is therefore not applicable at scales smaller than the dislocation spacing. In terms of crack tip analyses, this implies that CMSG plasticity predictions should be taken with caution for distances below 100 nm to the crack tip.

## 3. Material properties, loading conditions and finite element approach

To establish the crack behaviour under pure mode I and mode II, numerical calculations were performed covering a wide range of material properties by employing a compact tension shear (CTS) specimen, arguably the most popular testing configuration in mixed-mode studies (Fig. 1). A unit thickness $t$ is assumed, such that the stress intensity factors can be formulated as a function of the crack length $a$ and the specimen width $B$ as follows [33]:

$$K_{I} = \frac{P}{Bt}\sqrt{\pi a}\frac{\sin\beta}{1-\frac{a}{B}}\times\left[\frac{0.26-2.65\frac{a}{B-a}}{1+0.55\frac{a}{B-a}-0.08\left(\frac{a}{B-a}\right)^{2}}\right]^{1/2} \qquad (13)$$

$$K_{I} = \frac{P}{Bt}\sqrt{\pi a}\frac{\cos\beta}{1-\frac{a}{B}}\times\left[\frac{-0.23+1.4\frac{a}{B-a}}{1+0.67\frac{a}{B-a}+2.08\left(\frac{a}{B-a}\right)^{2}}\right]^{1/2} \qquad (14)$$

Pure mode I is obtained when the force $P$ is applied in a direction $\beta = 90°$, whereas pure mode II is achieved by applying the force $P$ in a direction $\beta = 0°$ (Fig. 1). The angle $\beta$ is measured from the plane of the original notch in the specimen. The crack faces remained traction-free in the two fracture modes considered.

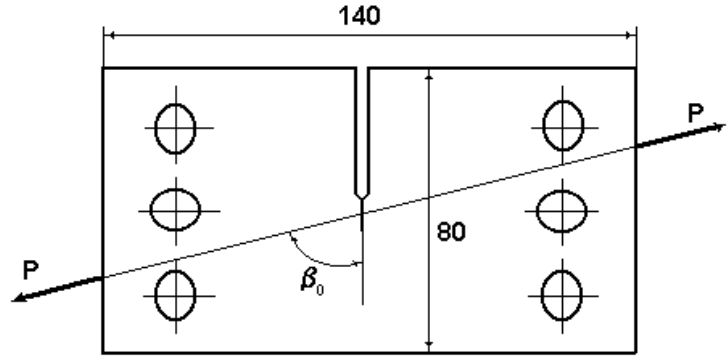

**Fig. 1.** Compact tension shear (CTS) specimen. Pure Mode I conditions are given by $\beta=90°$ while pure Mode II conditions are given by $\beta=0°$. Dimensions are given in mm.

One of the purposes of the present work is the estimation of stress-strain state (plane strain and plane stress) and material properties effects under both fracture modes. To this end, the material length scale was ranged from 1 to 20 μm, while the influence of the strain hardening exponent will be assessed by spanning the range $N=0.075$ to $N=0.4$ forming the basis for the parametric study.

To compare the mode I and mode II loading conditions, FE-computations are performed for the same remote elastic SIFs, namely, $K_1 = K_2 = 7.3$ MPa·m$^{0.5}$. The analysis was addressed to pure mode I and pure mode II fracture state, and the elastic SIFs were normalized as $\bar{K}_{1,2} = K_{1,2}/\sigma_y\sqrt{l}$. The material properties and loading conditions are summarized in Table 1.

**Table 1.** Material properties and loading conditions

| $E$ GPa | $\sigma_y$ MPa | $\nu$ | $N$ | $K_1$ MPa·m$^{0.5}$ | $K_2$ MPa·m$^{0.5}$ | $l$ μm | $\bar{K}_{1,2} = K_{1,2}/\sigma_y\sqrt{l}$ |
|---|---|---|---|---|---|---|---|
| 100 | 200 | 0.3 | 0.075-0.4 | 7.3 | 7.3 | 1 | 36.70 |
| | | | | | | 5 | 16.41 |
| | | | | | | 10 | 11.60 |
| | | | | | | 20 | 8.21 |

The CMSG plasticity theory described in Section 2 was implemented in the FE-code ANSYS [34] by means of a user material subroutine USERMAT. The implementation of mechanism-based SGP theories has been described in more detail in Refs. [35,36]. It should be emphasised that geometrical and material non-linearities are accounted for in the present implementation; rigid body rotations for the strains and stresses are carried out by means of the Hughes and Winget algorithm [37].

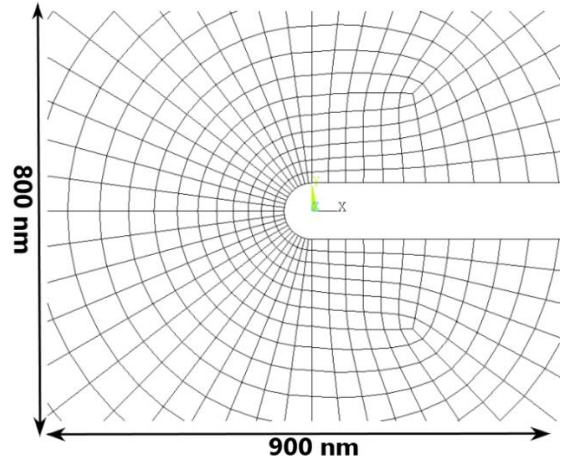

**Fig. 2.** FE mesh with the initial crack tip radius. The characteristic element size is approximately 6 nm.

We modelled a CTS specimen (Fig. 1) of in-plane dimensions 80 × 136 mm with initial relative crack length $a/w = 0.5$. The initial crack tip was assigned a radius of curvature $\rho = 60$ nm. The results presented were obtained with an initial notch root radius of approximately $10^{-5}$ times the characteristic crack dimension. Several different root radii were investigated and, as in the boundary layer calculations, we found that the stress and strain distributions were not dependent on the initial root radius when the crack tip was blunted beyond approximately three times the initial root radius. Also, the effect of the crack blunting radius was negligible at distances ahead of the crack tip where continuum models apply. Thus, the solutions were independent of the initial root radius and could be interpreted as those pertaining to an initially sharp crack. A mesh sensitivity analysis led to a characteristic element size of $h = 6$ nm that yields mesh-independent results [36]. A highly refined mesh was used near the crack tip (see Fig. 2). A typical mesh for the CTS specimen geometry has approximately 1,155,000 elements.

## 4. Crack tip stress behaviour

### 4.1. Hoop and effective stress radial distributions

In this section, we obtain the mode I and mode II plane strain and plane stress near-tip fields for cracks in elastic-plastic materials with strain gradient effects. Figure 3 illustrates the influence of the mode fracture type (I/II) and the plastic work hardening exponent on the hoop $\sigma_\theta/\sigma_Y$ and effective normalized von Mises stress $\sigma_e/\sigma_Y$ distribution ahead of the crack tip at $\theta = 0°$ in the CTS specimen.

The results depicted in Fig. 3(a,b) reveal several interesting effects. First, in agreement with the literature, incorporating the role of plastic strain gradients leads to a stress elevation within microns ahead of the crack tip. Far away from the crack tip, CMSG plasticity and conventional plasticity ($l = 0$, denoted as CPS) agree, but gradient effects become dominant within microns to the crack tip and this leads to a notable stress elevation. Second, strain gradient effects appear to be more significant in plane stress than in plane strain conditions. For example, see Fig. 3(a), the hoop stress at 0.2 μm to the crack tip is 5 times larger than the conventional plasticity solution in plane stress, while only 2.5 times larger under plane strain conditions. A higher susceptibility to the role of plastic strain gradients under in-plane stress conditions can be considered counter-intuitive, as the plastic zone size $R_p$ is smaller in plane strain, which would result in a larger $l/R_p$ ratio relative to the plane stress case, for a given applied load. However, as it will be shown later on, when comparing across different stress states a larger plastic region does not necessarily translate into a larger gradient-dominant region, where GNDs and plastic strains govern the mechanical response. The effective von Mises stress distribution is shown in Fig. 3(b). The plane stress results, for both conventional and strain gradient plasticity, lead to higher stress levels due to the presence of a non-zero out-of-plane stress $\sigma_{zz}$. Otherwise, the trends are similar to those observed with the hoop stress.

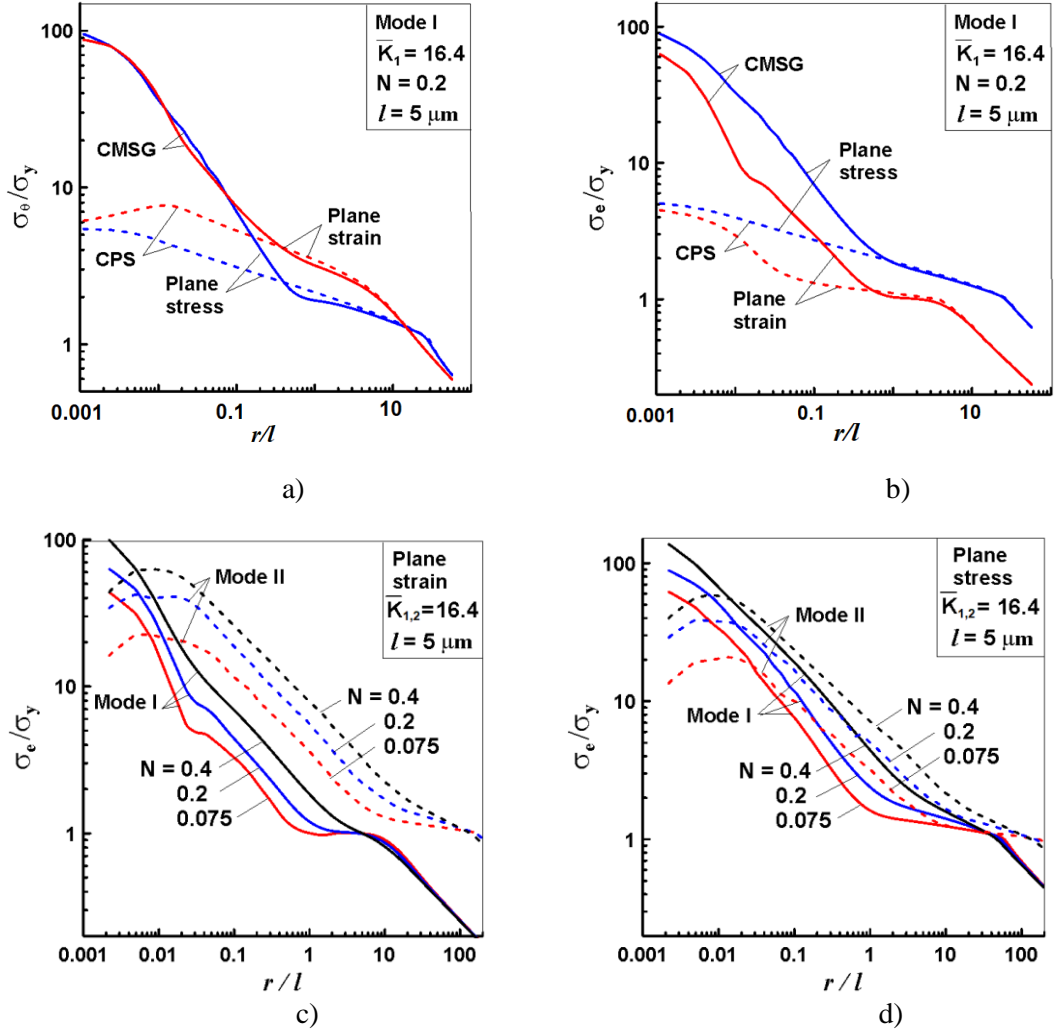

**Fig. 3.** Comparison of stress distributions ahead of the crack tip ($\theta = 0°$): (a) hoop $\sigma_\theta$ and (b) effective stress $\sigma_e$ for conventional (CPS) and gradient (CMSG) plasticity for mode I at $N = 0.2$; (c) plane strain and (d) plane stress results for different values of the strain hardening exponent $N$ and both mode I and mode II loading conditions. Material properties: $l = 5$ μm, $\sigma_y/E = 0.002$, $\nu = 0.3$, $\bar{K}_{1,2} = 16.4$.

In the Fig.3(c,d), the remote load $K_{1,2}/\sigma_Y\sqrt{l}$ of the applied SIFs is $\bar{K}_{1,2} = 16.4$, with the length parameter being $l = 5$ μm, and the plastic work hardening exponent values ranging as follows: $N = 0.075, 0.2$, and 0.4. The stress level increases with $N$ from 0.075 to 0.4, in agreement with expectations, and for both mode I and mode II, the asymptotic crack tip fields appear to exhibit some sensitivity to the strain hardening exponent. Under plane strain conditions, the normalised effective stress is larger for mode II loading, relative to mode I. However, under plane stress conditions, the mode I and mode II normalised effective stress distributions appear to be rather similar. The trends observed in Fig.3(c,d) persist for different levels of $K_{1,2}$ and for different values of the intrinsic parameter $l$.

Our computations confirm that large strain gradients exist near the crack tip, owing to the high stress and strain singularity. One of the important implications of the results depicted in Fig. 3(c,d) for CMSG plasticity is the sensitivity of the asymptotic stress singularity to mode I/II fracture and the strain hardening exponent $N$. As mentioned earlier, the subject of our analysis is the local material behaviour in the vicinity of a crack with an initial, finite blunting radius at the tip. In this region, under pure mode I loading conditions, the crack blunts along the direction of loading. However, under the pure mode II scenario, an opposite type of deformation develops at the upper and lower flanks of the crack tip; that is, sharpening in the upper region with blunting in the lower region. Creager and Paris [38], McMeeking [39], O'Dowd and Shih [40], and Mikkelsen and Goutianos [41] found that, except in the immediate vicinity of the blunted crack tip, the stress state coincides with the typical sharp crack tip stress field. Therefore, in the subsequent presentation of the singularity exponent, we consider the region within a distance of $5 \cdot 10^{-6} \leq r/a \leq 5 \cdot 10^{-4}$ from the crack tip, where the singularity of the effective stress distribution is uniform. The asymptotic crack tip singularity indexes obtained are presented in Table 1, for varying material length scale $l$ values, strain hardening exponent $N$, and both mode I and mode II conditions within plane stress and plane strain states.

**Table 2.** Values of crack tip singularity exponent $\gamma$

|  | Plane strain | | | | | |
|---|---|---|---|---|---|---|
|  | Mode I | | | Mode II | | |
|  | N = 0.075 | N = 0.2 | N = 0.4 | N = 0.075 | N = 0.2 | N = 0.4 |
| $l = 1$ μm | -0.654 | -0.657 | -0.666 | -0.542 | -0.568 | -0.572 |
| $l = 5$ μm | -0.651 | -0.665 | -0.632 | -0.526 | -0.547 | -0.532 |
| $l = 10$ μm | -0.658 | -0.666 | -0.608 | -0.513 | -0.536 | -0.514 |
| $l = 20$ μm | -0.658 | -0.661 | -0.593 | -0.500 | -0.527 | -0.502 |
|  | Plane stress | | | | | |
|  | Mode I | | | Mode II | | |
|  | N = 0.075 | N = 0.2 | N = 0.4 | N = 0.075 | N = 0.2 | N = 0.4 |
| $l = 1$ μm | -0.621 | -0.617 | -0.617 | -0.525 | -0.523 | -0.544 |
| $l = 5$ μm | -0.648 | -0.645 | -0.597 | -0.506 | -0.512 | -0.517 |
| $l = 10$ μm | -0.672 | -0.642 | -0.582 | -0.501 | -0.521 | -0.510 |
| $l = 20$ μm | -0.659 | -0.639 | -0.543 | -0.501 | -0.514 | -0.504 |

Several interesting features can be inferred from the results shown in Table 2. First, the singularity values for the mode I plane strain and plane stress are larger than those obtained for pure mode II at an equivalent stress-strain state. Moreover, the singularity exponents are in all cases (including mode II) larger than the classical linear elastic result $r^{-1/2}$. Thus, SGP models based on Taylor's dislocation theory appear to predict a stronger singularity than linear elastic solids for both plane strain and plane stress under mode I and mode II loading conditions. Shi et al. [44] also found a higher stress singularity than that of linear elasticity in the case of the higher-order MSG plasticity model under mode I conditions; they concluded that the stresses scale as $r^{-2/3}$ in the vicinity of the crack tip. Second, the magnitude of the stress singularity exhibits some sensitivity to the intrinsic material length parameter $l$, with differences becoming more noticeable for larger strain hardening values. For the case of $N = 0.4$, larger values of $l$ appear to bring the solution closer to the linear elastic result. The singularity exponents estimated differ from those reported in the literature. They are similar to those reported for MSG plasticity [42,43] but exhibit a sensitivity to the strain hardening exponent. And are higher than those reported for higher order phenomenological strain gradient plasticity models [24,26,44,45].

*4.2. Stress triaxiality angular distributions*

The role of stress triaxiality conditions on the fracture behaviour of elastic-plastic materials has always been a topic of significant interest [39,46-49]. A local parameter of stress triaxiality and crack tip constraint was proposed by Henry and Luxmoore [50] as follows:

$$h = \frac{\sigma_{kk}}{\left(3\sqrt{S_{ij}S_{ij}}\right)} \qquad (15)$$

where $\sigma_{kk}$ and $S_{ij}$ are the hydrostatic and deviatoric stresses. The angular distribution of $h$ is computed here using CMSG plasticity as a function of $N$, the distance ahead of the crack $r/l$ and the fracture mode – results are shown in Fig. 6. Both plane stress and plane strain results are shown and the polar angle $\theta$ is measured from the symmetry line in front of the crack tip.

Consider first the mode I case. The results shown in Fig. 4a show that the maximum triaxiality levels in the plane strain mode I near-tip fields in materials with strain gradient effects can be up to 5 times larger than those relevant to plane stress conditions, for $N = 0.2$. As shown in Fig. 4b, these differences increase with smaller strain hardening exponents. The constraint effect ahead of a crack tip at the polar angle $\theta = 0°$ in the CTS specimen under mode I plane strain conditions increases with a variation in the hardening exponent from 0.075 to 0.4. A strong sensitivity of the stress triaxiality parameter $h$ on the distance from the crack tip and the strain hardening exponent is observed. The sequence of curves in Fig. 4a as a function of crack tip distance demonstrates the typical distribution due to the blunting of the crack tip. Now, let us turn our attention to the mode II results, Figs. 4b and 4d. It is observed that the plane stress and plane strain results in Figs. 4b and 4d are much closer to each other than for pure mode I. This statement holds true for various distances from the crack tip $r/l$ and work hardening exponents $N$. The distributions of the triaxiality parameter for mode II are almost insensitive to variations in the work hardening and crack tip distance. The reduction in triaxiality observed for the mode II conditions is accompanied by an increase in plastic deformations, which in turns contributes to transitioning to a more ductile fracture scenario. However, in the high-constraint case, produced by the pure mode I plane strain conditions, the triaxial stress increases significantly while the plastic strain in the $\theta = 0°$ case remains relatively low. These conditions favour cleavage fracture. In general, the stress triaxiality trends are similar to those reported for conventional plasticity, as plastic strain gradients contribute to both the effective stress and the hydrostatic stress.

In contrast to mode I, under the pure shear condition the triaxial stress has two maxima, which are located at angular coordinates of $\theta = \pm(65–75)°$ and almost coincide for both plane strain and plane stress states (Figs. 4b and 4d). Figure 4 depicts the triaxial stress variation as a function of the material plastic properties under two fracture mode conditions. In agreement with expectations, the triaxial stress level decreases as the mode changes from I to II. This trend has a significant effect on the distribution of dislocation densities, as the dislocation behaviour is likely to be influenced by stress triaxiality. Therefore, the distributions of the stress triaxiality parameter shown in Fig. 4 are further used to analyse the dislocation density behaviour at the crack tip in accordance with the present theory of Taylor-based gradient plasticity.

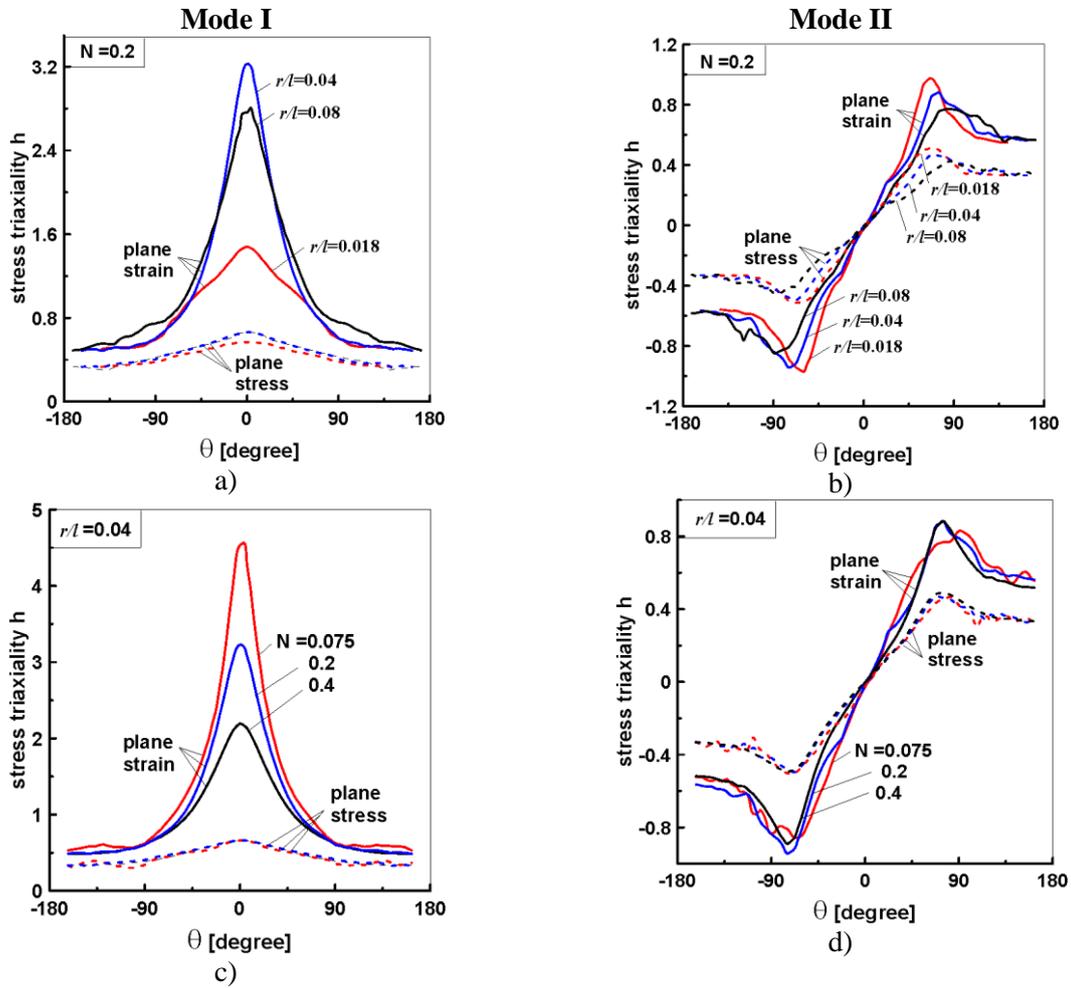

**Fig. 4.** Mode I and mode II angular distributions of the stress triaxiality at $\bar{K}_{1,2} = 16.4$ as a function of different (a,b) locations ahead of the crack tip (for $N = 0.2$) and (c,d) strain hardening exponent (for $r/l = 0.04$). Material properties: $\sigma_y/E = 0.002$, $\nu = 0.3$ and $l = 5$ μm.

## 5. Plastic stress intensity factors for mode I/II behaviour

One of the main elements of the structure of stress, strain, and displacement fields at the crack tip are scale factors in the form of amplitude coefficients or stress intensity factors (SIFs). These factors have the physical meaning of measures of near-field loading close to the crack tip and are generally dependent on the applied load, the cracked body configuration, and the elastic-plastic material properties. There are limited studies in the literature on stress intensity factors in SGP, and these studies are limited to a few particular cases and do not have sufficient generalization. First, to formulate the problem for the crack tip fields, Xia and Hutchinson [26] determined, similar to the classical HRR solution, the plastic SIF, and derived a new amplitude factor, which was related to the dimensionless angular variations for dominant stress components. Later, Huang et al. [27] proposed several equations for the elastic and plastic amplitude factors for the asymptotic crack tip fields in materials with strain gradient effects. It is stated that under the condition of pure elasticity, as well as elastic-plastic strain gradient materials with a separated or integrated law of hardening, the dominant stress and coupled stress fields in mode I are governed by two independent parameters. Recently, Shlyannikov et al. [36] presented numerical and analytical formulations for the plastic SIFs, which are measures of crack tip stress amplitude and are applicable in the domain of validity of CMSG plasticity.

In this section, following Shlyannikov et al. [36], we adopt the numerical formulation for the amplitude $A_P^{FEM}(r,\theta)$ and plastic stress intensity $K_P^{FEM}$ factors for CMSG plasticity, which are given by:

$$\bar{\sigma}_e^{FEM}(r,\theta) = K_P^{FEM} \bar{r}^\gamma \hat{\sigma}_e^{FEM}(r,\theta) \tag{16}$$

$$A_P^{FEM}(r,\theta) = \bar{\sigma}_{ij}^{FEM}(r,\theta) / \hat{\sigma}_{ij}^{FEM}(r,\theta) \tag{17}$$

$$K_P^{FEM} = A_P^{FEM} / \bar{r}^\gamma . \tag{18}$$

where $\bar{r} = r/l$ is the nondimensional distance to the crack tip and $\gamma$ is the power of the stress singularity. In Eqs. (16)-(17), the angular distributions of the stress component $\hat{\sigma}_{ij}^{FEM}(r,\theta)$ are normalised, such that $\hat{\sigma}_{e,\max}^{FEM} = (3/2 S_{ij}^{FEM} S_{ij}^{FEM})_{\max}^{1/2} = 1$ and $\bar{\sigma}_{ij}^{FEM} = \sigma_{ij}^{FEM} / \sigma_Y$. For all the considered combinations of the intrinsic material length parameter $l$ and the plastic work hardening exponent $N$ under plane strain and plane stress states and mode I and mode II loading conditions, the values of the power of stress singularity $\gamma$ are listed in Table 2. The constitutive laws for the CMSG model are some of the simplest generalisations of the $J_2$ flow theory of plasticity that include strain gradient effects. Our calculations indicate that stresses have a variable singularity near the crack tip and are governed by the plastic stress intensity factor $K_P^{FEM}$ in accordance with Eq. (18), which is determined from the condition of matching the numerical solution. As in conventional plasticity, we demonstrate that a plastic stress intensity factor $K_P$ can be defined, which exhibits a constant value within the region of gradient dominance for a fixed value of $N$.

The calculation of the plastic SIF $K_P^{FEM}$ reveals an almost uniform magnitude over the gradient dominated zone ahead of the crack tip. Consequently, we take the converged value of $K_P^{FEM}$ and plot its radial distributions in Fig. 5 under the mode I and II plane stress and plane strain states for different values of the intrinsic material length $l$ and the plastic work hardening exponent $N$. The results show the sensitivity of the employed parameter of fracture resistance $K_P^{FEM}$ in the assessment of the coupled effects of parameters $l$ and *N*. A monotonic decrease of the plastic SIFs is observed when increasing the material length parameter from 1 to 20 μm. The smaller values of the plastic SIF obtained with increasing *l* are a consequence of the larger influence of plastic strain gradients, which increase plastic flow resistance. Independently of the magnitude of the CMSG plasticity length scale, a higher level of the plastic SIF can be observed for mode II case, relative to mode I conditions – see Fig. 5a. This can be rationalised with the triaxiality results observed before – plastic dissipation is notably higher under mode II fracture conditions.

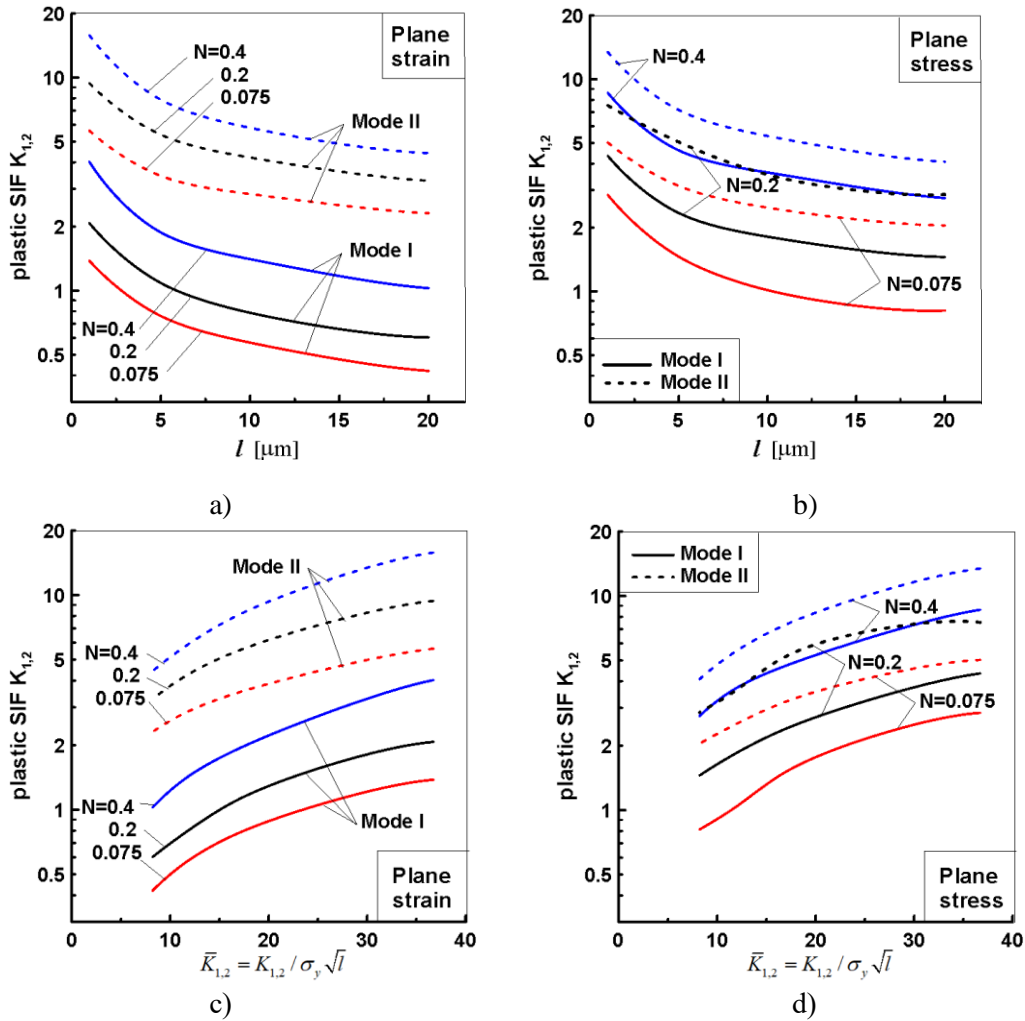

**Fig. 5.** Plane strain and plane stress plastic SIF behaviour as a function of (a,b) the material length parameter *l* and (c,d) the remote normalized elastic SIFs for different fracture modes and strain hardening exponents *N*.

An important conclusion regarding the numerical results depicted in Figs. 5c and 5d is that for the same values of the remote elastic SIFs $\bar{K}_{1,2}$, the magnitude of the plastic SIFs $K_P^{FEM}$ monotonically increases with the work hardening exponent *N*. It follows from Fig. 5c and 5d that the plastic SIFs increase more gradually than the elastic ones owing to the redistribution of stresses in the plastic region of the crack tip. With respect to the classical elastic SIFs, the new plastic factors $K_P^{FEM}$ differ in the plane strain and plane stress states and exhibit the property of being sensitive to the coupling effect of the material scale of length and the work hardening exponent, which is important from the perspective of practical applications.

## 6. GND and SSD density distributions

We proceed to analyse the GND and SSD density behaviour as predicted by Taylor's model and CMSG plasticity, see Eqs. (3)–(4). It is well known that the plastic deformation of metals is typically accommodated by dislocation motion. The progress of slipping dislocations through the material controls the mechanical properties, such as yield strength, work hardening, and crack growth resistance. Both GNDs and SSDs are related to work hardening; however, their nucleation and local contributions are slightly different: GNDs arise owing to plastic strain gradients (lattice curvature), whereas SSDs are thought to evolve as a function of the total plastic strain. As in previous sections, we aim at assessing the coupling of material properties and SGP effects. To this end, a wide range of values for the plastic work hardening exponent $N$ and the intrinsic material length parameter $l$ are used in the computations. In particular, $N$ is varied between 0.075 and 0.4, while the material length scale is varied between 1 and 20 μm. Changing the stress-strain state by considering both plane strain and plane stress within pure mode I and mode II fracture conditions enables us to further expand the scenarios where gradient effects and dislocation behaviour will be assessed. The sensitivity of the results to the remote load is explored and the quantity $K_{1,2}/\sigma_y\sqrt{l}$ is varied across a wide range of 8.21–36.7. Our numerical results in this section on the distribution and evolution dislocation densities are presented as radial and angular fields of both GND and SSD densities, including colour illustrations of the corresponding contours in the vicinity of the crack tip.

### 6.1. Radial dislocation density distributions

GND density distributions obtained from the first-order mechanism-based SGP analysis for mode I and mode II are compared under plane strain and plane stress in Figs. 6a and 6b, respectively. Here, $r$ denotes the radial distance from the tip, which is normalized by the intrinsic material length parameter $l$. The applied mode I and mode II SIFs are $\bar{K}_{1,2} = 16.4$ and the intrinsic material length equals $l = 5$ μm. The plastic work hardening effect is significant for $r/l < 1$, because the GND density increases by nearly an order of magnitude when $N$ varies from 0.4 to 0.075. In mode I, the stress dislocation density is elevated slightly above the plane stress distributions (Fig. 6a), and they are closer to each other for the mode II distributions (Fig. 6b).

The radial variations in the densities $\rho_G$ and $\rho_S$ of GNDs and SSDs along $\theta = 0°$ are shown in Figs. 6c and 6d as a function of the intrinsic material length parameter $l$. The radial distance of the material in the undeformed state is normalised by the crack length $a$ because, in this case, we consider the variation of the material scale factor $l$. Here, $\rho_S$ and $\rho_G$ are related to the uniaxial stress–plastic strain relation and the effective plastic strain gradient, as $\rho_G = \bar{r}\eta^P/b$ and $\rho_S = \left[\sigma_{ref}f(\varepsilon^P)/M\alpha\mu b\right]^2$, according to Eq. (3) and (4), respectively. From the results in Figs. 6c and 6d, it can be seen that under both mode I and mode II plane strain conditions, the total dislocation density decreases when the intrinsic material length parameter $l$ changes from 1 to 20 μm. The smaller the length scale the smaller the capacity of the material to harden with plastic strain gradients, and thus the larger the degree of plastic deformation. Recall from Eq. (8) that strain gradient hardening becomes less important as the term associated with the equivalent plastic strain becomes large relative to the gradient term $l\eta^p$.

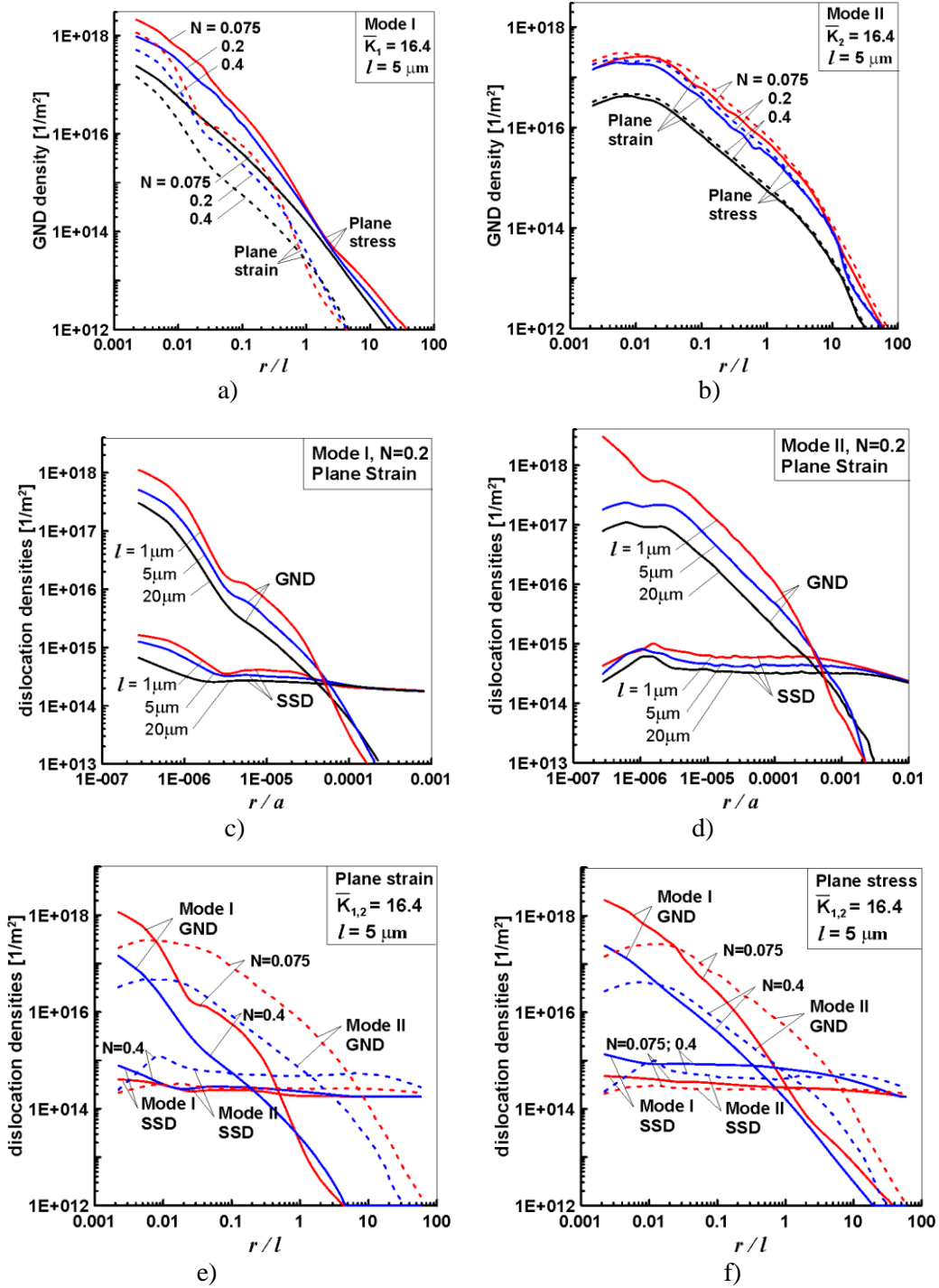

**Fig. 6.** Comparison of plane strain and plane stress GND and SSD density radial distributions for different loading conditions and plastic material properties: $\sigma_y/E = 0.002$, $\nu = 0.3$, (a,b,e,f) $l = 5$ μm, (c,d) $N = 0.2$.

As shown in Figs. 6c and 6d, the distributions of the SSD and GND densities are notably different. The GND density $\rho_G$ is large around the crack tip, but rapidly decreases away from it. On the contrary, the SSD density $\rho_S$ is not as large as that of $\rho_G$ around the crack tip, but decreases much more slowly than $\rho_G$ away from it. Both $\rho_S$ and $\rho_G$ are above $10^{14}$ m$^{-2}$ near the crack tip, but the magnitude of the GND density can be roughly 3 orders of magnitude larger than the SSD density in the vicinity of the crack tip. The results emphasise the role of GNDs in governing the mechanical behaviour of solids at very short distances ahead of the crack tip. Figures 6c and 6d assess the influence of varying the strain hardening exponent $N$. It can be observed that, the smaller the strain hardening exponent, the larger the GND density, with very little sensitivity observed for the SSD density. Thus, results suggest that a material with low conventional hardening capacity will lead to a larger accumulation of GNDs. Comparisons of plane strain and plane stress GND and SSD density radial distributions for different fracture modes I and II are shown in Figs. 6e and 6f, respectively. The results are expressed as a function of the normalized crack tip distances $r/l$ for two different values of the strain hardening exponent $N = 0.075$ and $0.4$, with the applied mode I and mode II SIF of $\bar{K}_{1,2} = 16.4$, and an intrinsic material length equal to $l = 5$ μm. The fracture mode effect on both GND and SSD density behaviour is significant, particularly under plane strain conditions. In addition, in all the cases considered, the GND and SSD dislocation densities for mode II are higher than that for mode I if $r/l > 0.05$, but a change in trend is observed closer to the crack tip.

### *6.2. Contour plots for dislocation densities*

The colour contour plots of the GND and SSD densities are shown in Fig. 7 for plane strain and plane stress mode I and mode II loading conditions. The unit of dislocation density is m$^{-2}$ and, for illustrative purposes, the GND $\rho_G$ and SSD $\rho_S$ density contours in Fig. 7 are selected in the range of $1\cdot10^{15}$–$5\cdot10^{16}$ m$^{-2}$. The plots are shown for the strain hardening exponent $N = 0.2$, the applied mode I and mode II SIF $\bar{K}_{1,2} = 16.4$, and the material length $l = 5$ μm. Qualitatively, the contours of the GND density in Fig. 7 under plane strain and plane stress show a similar shape to that of the plastic zone surrounding a crack. For the particular case of plane strain and mode I loading, our results are in good agreement with the GND distributions obtained by Qu et al. (2004). The behaviour of the SSD density for mode II under the plane strain and plane stress states in Fig. 7 reveals that a large area with high SSD density values is located behind the crack front.

More detailed information on the dislocation density fields is given by the GND and SSD contours in Fig. 8, which are shown on the scale of the radius of curvature of the crack tip. The comparison of the contours for the GND and SSD densities reveals that in the nearest crack tip region, the GND density $\rho_G$ fields are more uniform, with a gradual decrease in intensity with increasing distance from the crack tip. The maximum values of the GND dislocation density on the considered scale reaches a magnitude of $1\cdot10^{18}$ m$^{-2}$. The GND density dislocation distribution in Fig. 8 under mode I plane strain and plane stress uniformly covers the blunted crack tip. Under the pure mode II condition, the results show that the GNDs are concentrated on the upper blunt crack tip side and the lower sharp corner. When we relate this tendency to the stress triaxiality distributions shown in Figs. 4b and 4d, we can infer that the higher values of stress triaxiality occur at the same upper and lower crack tip regions with a high GND density. Small quantitative differences in the GND density distribution tendency between these two regions can be observed, such that the maximum value of $\rho_G$ is reached in the upper blunt region. Assuming that fracture initiates at the location at which the dislocation density is high, we can infer that under the pure mode II condition, the initiation of crack growth will occur at the upper region of the crack tip. This assumption is in good agreement with experimental results obtained for different materials [51,52]. The concentration of SSD densities in these regions is less pronounced. Thus, high stress triaxiality locations tend to have high dislocation densities associated with them.

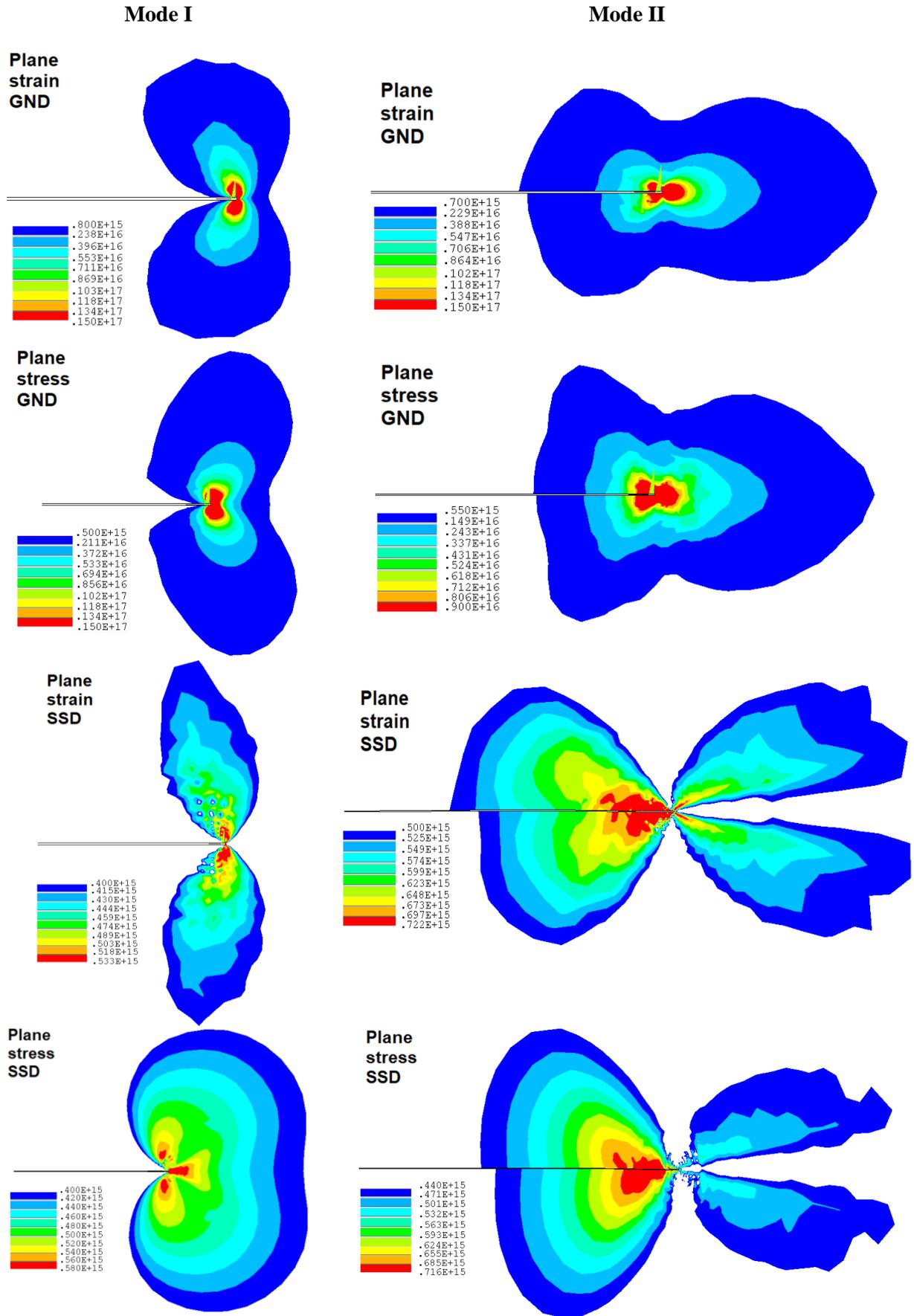

**Fig. 7.** Contour plots of GND and SSD densities (in m$^{-2}$) for plane strain and plane stress under mode I and mode II loading conditions. Dislocation densities become significant close to the crack tip. Material properties: $\sigma_y/E = 0.002$, $\nu = 0.3$, $l = 5$ μm, and $N = 0.2$.

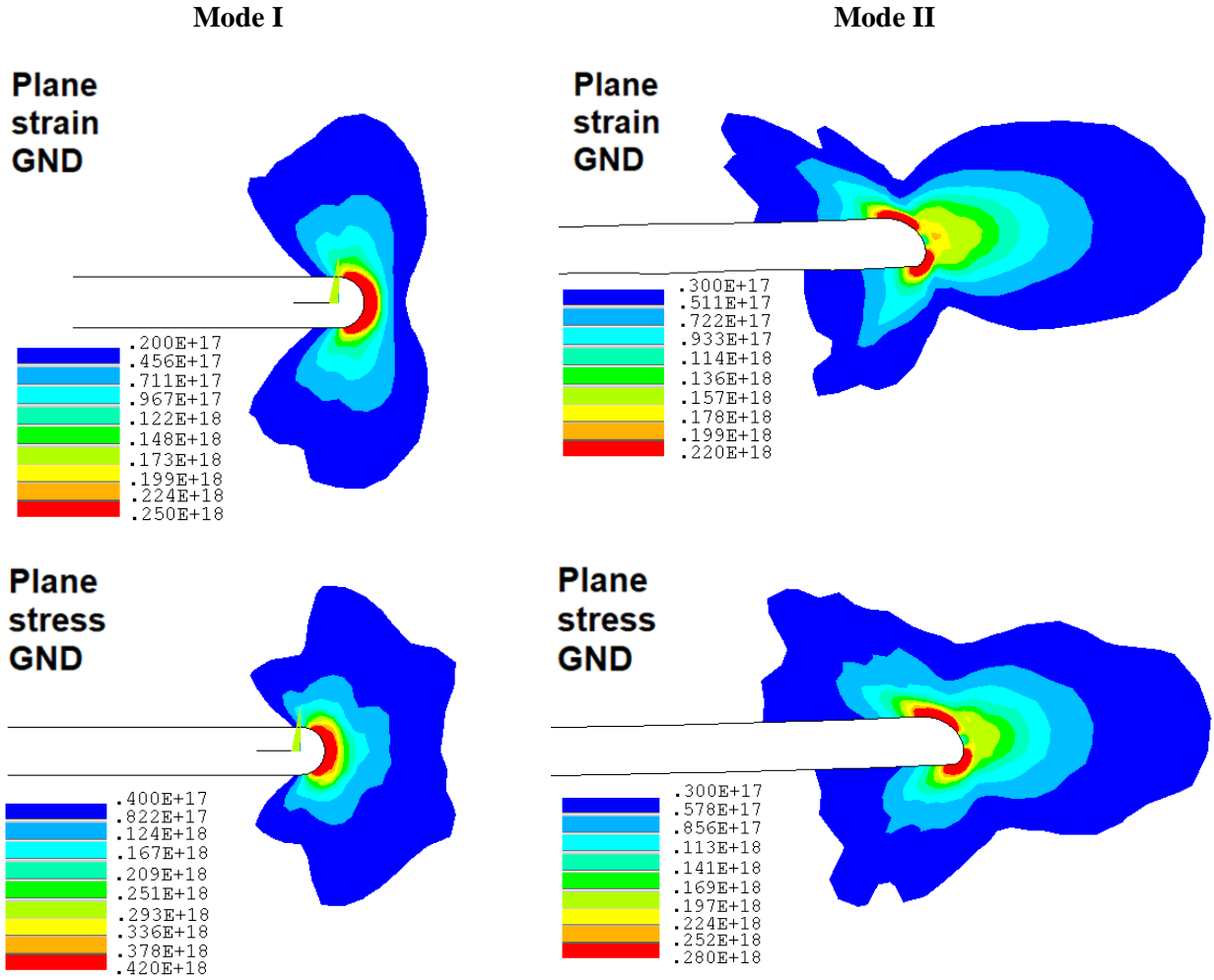

**Fig. 8.** Near crack tip contour plots of GND and SSD densities (in m$^{-2}$) for plane strain and plane stress under mode I and mode II loading conditions. Material properties: $\sigma_y/E = 0.002$, $\nu = 0.3$, $l = 5$ μm, and $N = 0.2$.

*6.3. Angular dislocation density distributions*

The angular distribution of the GND density $\rho_G$ is shown in Fig. 9 under various loading conditions and stress-strain states, where $\rho_G \cos\theta$ and $\rho_G \sin\theta$ constitute the axes (rectangular Cartesian coordinates). Results are shown for intrinsic length scales of $l = 5$ and 20 μm, applied mode I and mode II SIFs of $\bar{K}_{1,2} = 16.4$ and 8.2, strain hardening exponent of $N = 0.2$, and normalized crack tip distances of $r/a = 5 \cdot 10^{-6}$, $1 \cdot 10^{-5}$, and $1.5 \cdot 10^{-5}$. These selected $r$ values cover the region closest to the crack tip blunting area and the strain gradient dominated zone with a uniform stress singularity index. As before, the following material parameters $\sigma_y/E = 0.002$ and $\nu = 0.3$ are used. In agreement with expectations, the GND region expands with decreasing crack tip distance $r/a$; the results are consistent with those depicted in Fig. 7. For all the considered distances to the crack tip, the regions of dislocation densities $\rho_G$ for modes I and II are different, but their relative size is similar. It is also seen that the intrinsic material length parameter $l$ has a moderate effect on the qualitative shape of the GND density angular distributions under both plane strain and plane stress states.

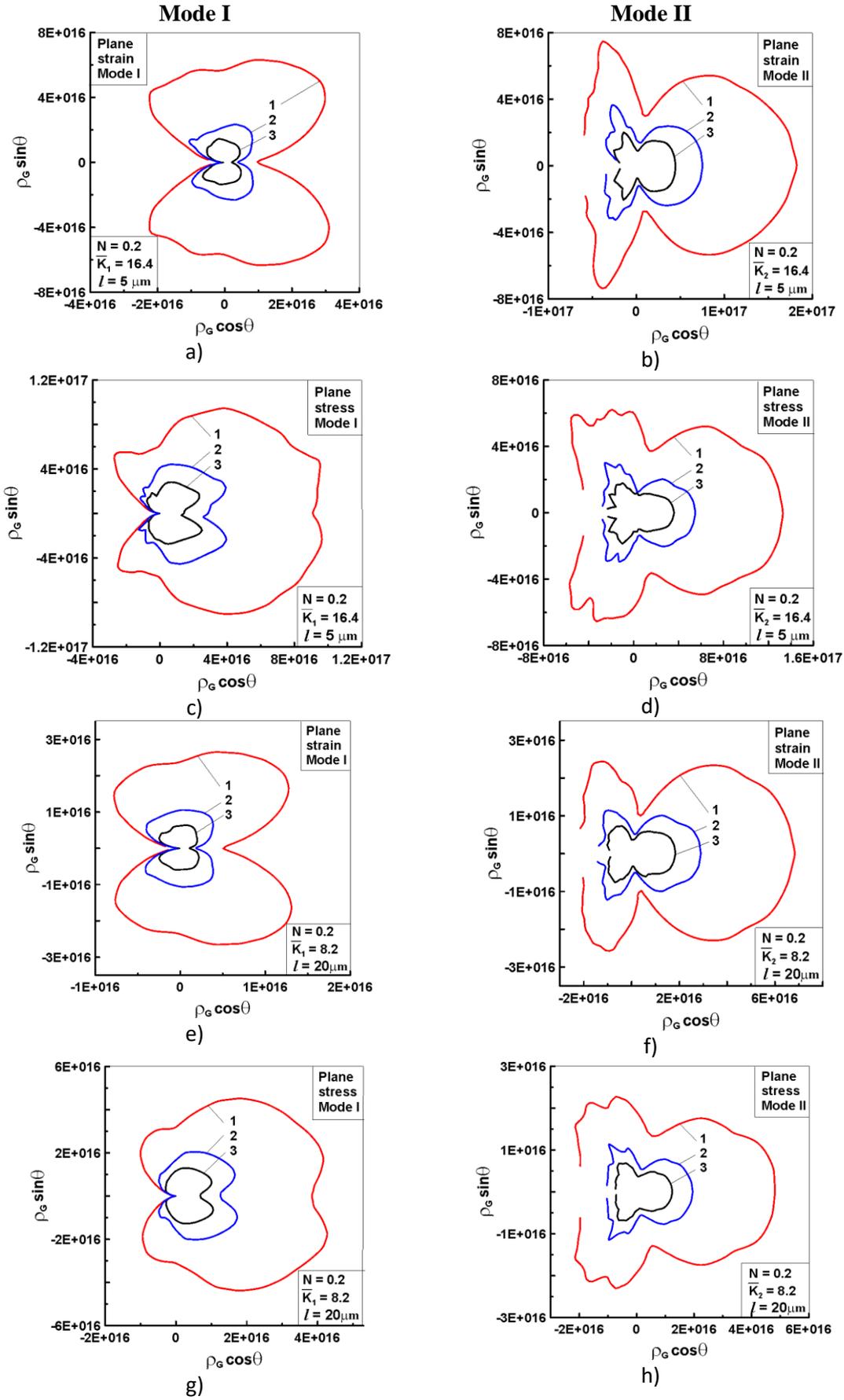

**Fig. 9.** Plane strain and plane stress angular distributions of GND density for mode I and mode II as a function of the material length parameter (a-d) $l = 5$ μm and (e-h) $l = 20$ μm and locations ahead of the crack tip: 1- $r/a=5\cdot10^{-6}$, 2- $r/a=1\cdot10^{-5}$, 3- $r/a=1.5\cdot10^{-5}$. Material properties: $\sigma_y/E = 0.002$, $\nu = 0.3$ and $N=0.2$.

A more detailed analysis of the influence of the plastic properties of the material on the angular distributions of the GND density is presented in Fig. 10. The distributions of the GND densities are presented for the distance $r/l = 0.08$, which is located in the region where the stress distribution exhibits a rather uniform singularity. The plastic work hardening exponent $N$ changes from 0.075 to 0.4, the remote normalized mode I and mode II SIF is $\bar{K}_{1,2} = 16.4$, and the intrinsic material length is $l = 5$ μm.

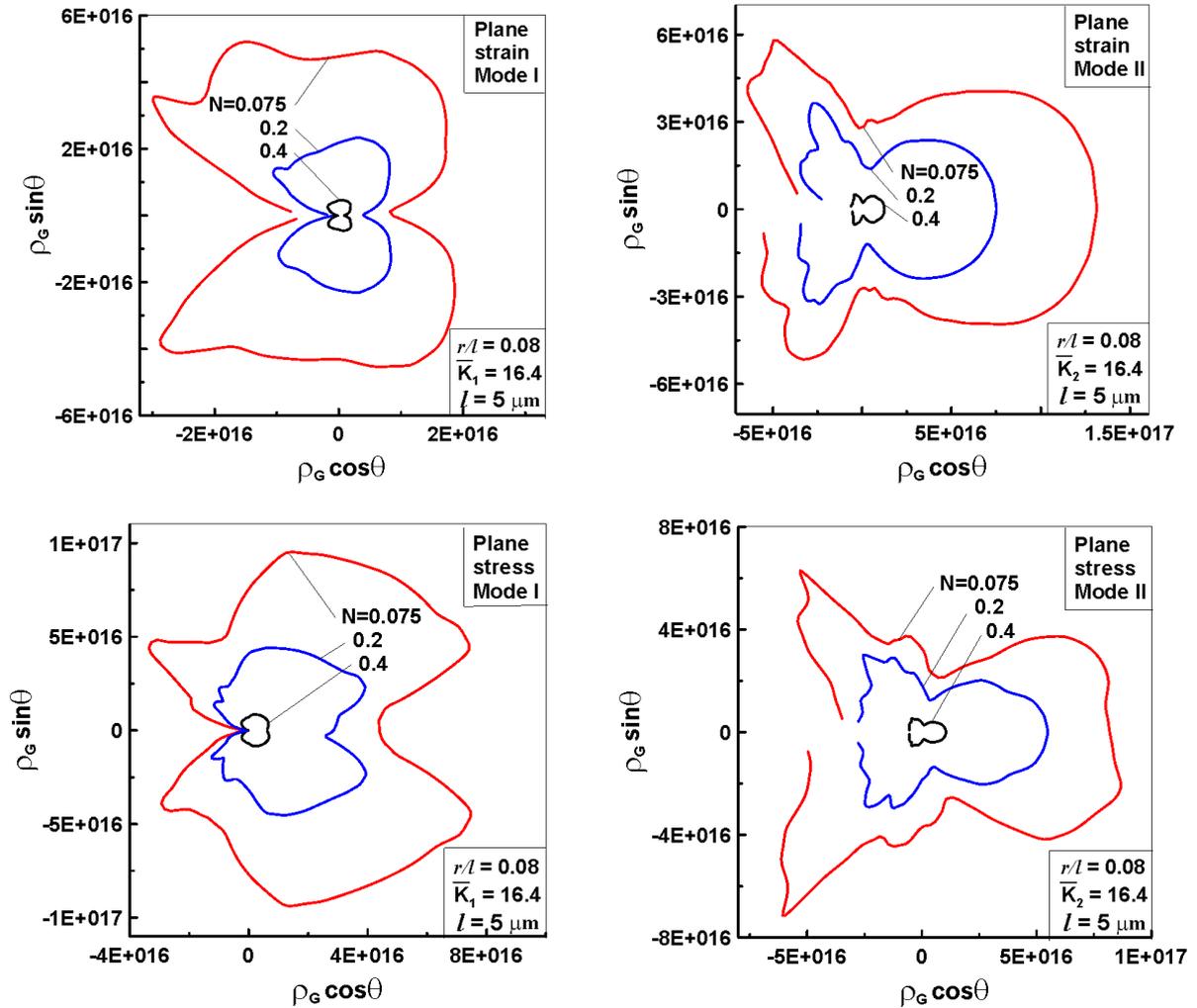

**Fig. 10.** Plane strain and plane stress angular distributions of GND density for mode I and mode II as a function of the work hardening exponent $N$ at a location ahead of the crack tip $r/l = 0.08$. Material properties: $\sigma_y/E = 0.002$, $\nu = 0.3$, and $l = 5$ μm.

The pattern of contours for the GND density $\rho_G$ plotted in Fig. 10 appears to depend strongly on the plastic work hardening exponent $N$, as well as on the fracture mode I/II. Varying of $N$ from a nearly elastic state with $N = 0.4$ to extended plasticity at $N = 0.075$ causes the GND density contours to increase in size under both plane strain and plane stress conditions. The dislocation density contours for $N = 0.075$ are approximately five times larger than those at $N = 0.4$. A similar behaviour is observed for a wide range of values of the material length parameter $l$.

We proceed to investigate the angular distribution of the SSD density and to compare it with the results obtained for the GND density for different strain hardening exponents and stress states – results are shown in Fig. 11. These distributions of the dislocation densities are given for the distance $r/l = 0.72$, plastic work hardening exponent values of $N = 0.2$ and $0.4$, remote normalized mode I SIF of $\bar{K}_1 = 16.4$, and intrinsic material length parameter of $l = 5$ μm.

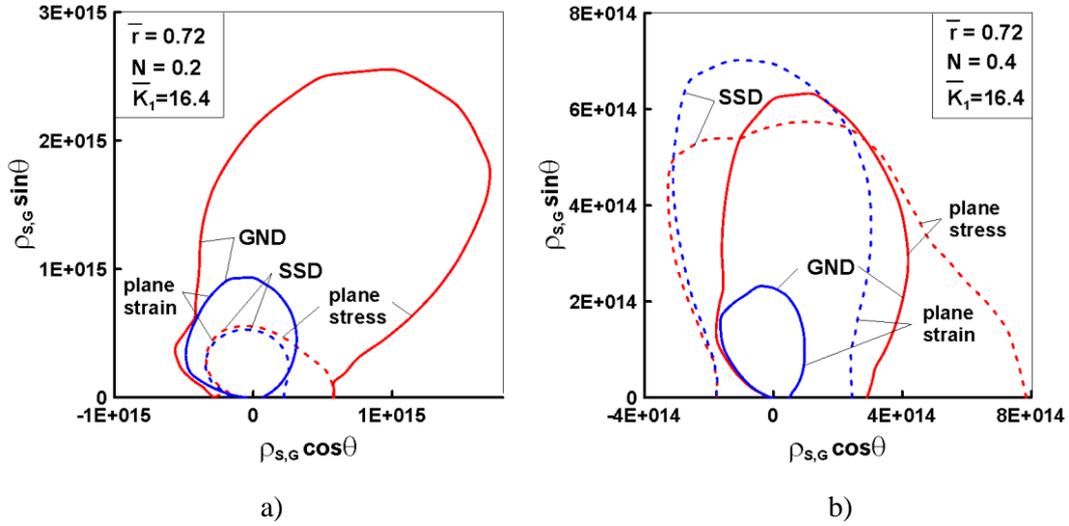

**Fig. 11.** Comparison of plane strain and plane stress angular distributions of GND and SSD densities for mode I as a function of the work hardening exponent $N$ at a location ahead of the crack tip $r/l = 0.72$. Material properties: $\sigma_y/E = 0.002$, $\nu = 0.3$, and $l = 5$ μm.

From the data presented in Fig. 11, it follows that for the same distance to the crack tip $r/l = 0.72$, opposite situations can occur in the absolute values of the GND and SSD dislocation densities, depending on the plastic properties of the material. Thus, for a sufficiently ductile material with $N = 0.2$ (Fig. 11a), the GND values prevail over the SSD values under the plane strain and plane stress states, and the maximum dislocation density is located at different angular coordinates that do not coincide with the line $\theta = 0°$ ahead of the crack tip. For a material with a much higher strain hardening exponent $N = 0.4$ (Fig. 11b), higher values are observed for the SSD density in comparison to those for the GND density under the same loading conditions.

## 7. Discussion

The present analysis of the radial (Fig. 6) and angular (Figs. 9–11) density distributions of GNDs and SSDs, $\rho_G$ and $\rho_S$, is fundamental to understand the effects of plastic material properties and strain gradients on crack tip mechanics. Our results imply that at small distances to the crack tip $r/l$, the gradient term $l\eta^P$ suppresses the effect of the plastic properties of the material. As the distance to the crack tip increases, the contributions of $\rho_G$ and $\rho_S$ become comparable, and for $r/l > 0.7$, the influence of plastic properties, characterised by the value of the plastic work hardening exponent $N$ in the first term in Eq. (8), $f^2(\varepsilon^P)$, prevails over the gradient effects.

Jiang et al. [16] used EBSD and HR-DIC to measure dislocation densities and showed that the range of variation in the GND density values is on the order of $10^{14}$–$10^{16}$ m$^{-2}$. A higher resolution is limited by the capabilities of the measurement tools. Taking into account the results of our numerical calculations, wherein the GND density varies in the range of $10^{13}$–$10^{18}$ m$^{-2}$, it can be assumed that the given order of experimental values of the GND density $10^{14}$–$10^{16}$ m$^{-2}$ corresponds to a distance to the crack tip at which the contributions of the GND and SSD densities are comparable to each other. Therefore, problems may arise in the interpretation of experimental data, also because of the density of the SSD being difficult to determine and identify. The results in Fig. 11 indicate that, depending on the crack tip distance, there can be opposite assessments of the maxima of the GND and SSD densities.

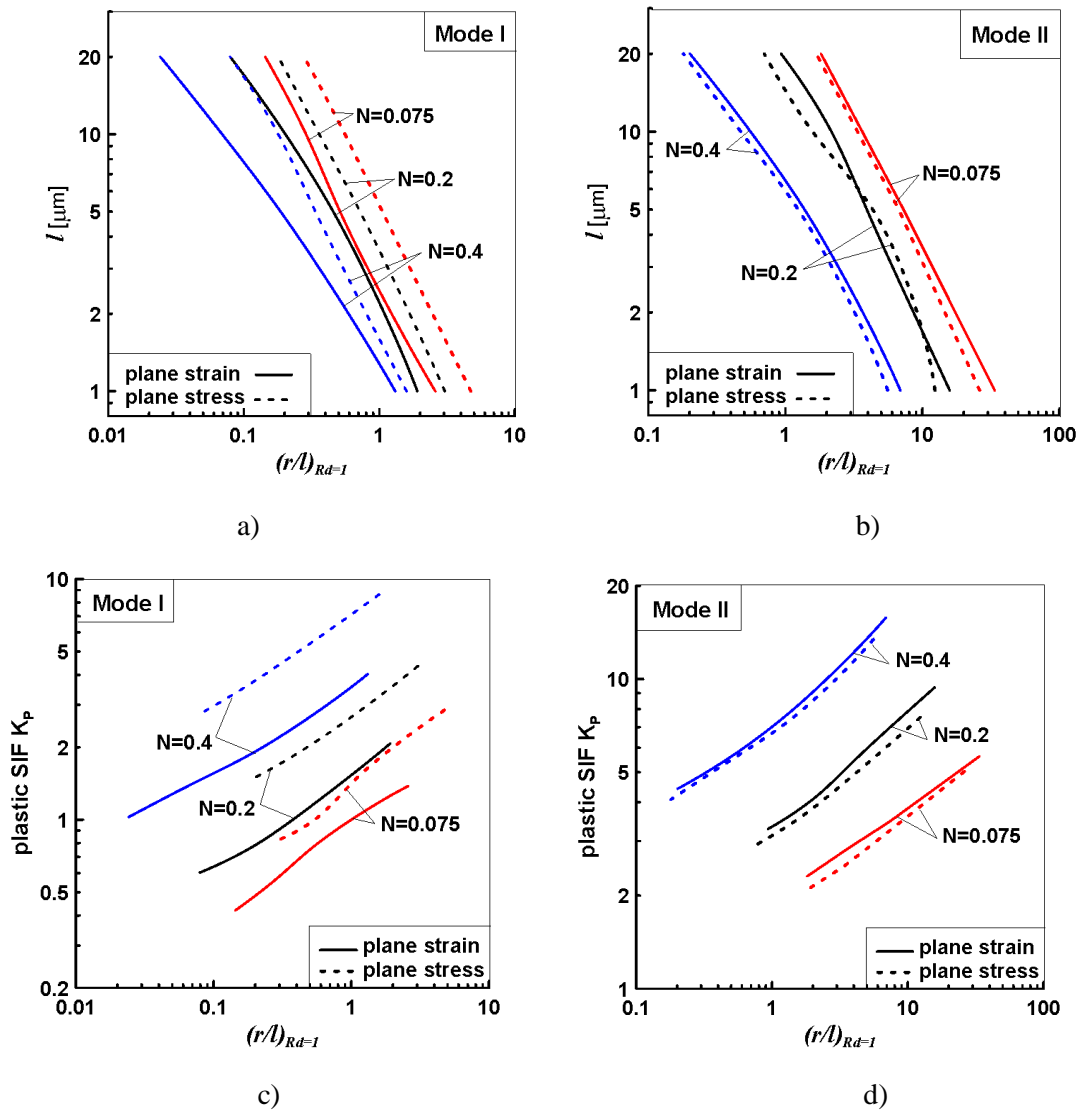

**Fig. 12.** Material length parameter (a,b) and plastic SIF (c,d) versus coordinate of the equilibrium state of dislocation densities for plane strain and plane stress under mode I and mode II as a function of the work hardening exponent *N*. Material properties: $\sigma_y/E = 0.002$, $\nu = 0.3$.

From an experimental perspective, it is useful to establish the radial coordinate of the equilibrium state of the dislocation densities. To this end, we introduce the density locations ratio in the following form:

$$R_d = \frac{\rho_G}{\rho_S} = \bar{r}\frac{\eta^P}{b}\left[\frac{M\alpha\mu b}{\sigma_{ref} f\left(\varepsilon^P\right)}\right]^2 = 0.988\frac{\eta^P l}{\left(\varepsilon^P + \sigma_y/E\right)^{2N}} \qquad (19)$$

Equality of this ratio $R_d = 1$ sets the distance ahead of the crack tip where an equilibrium state of dislocation densities is attained. Figure 12 shows the relation between the coordinates of the equilibrium state of dislocation densities $R_d = 1$ and the intrinsic material length as well as the plastic SIF $K_P^{FEM}$, as a function of the work hardening exponent $N$. Here, the dimensionless coordinate $(r/l)|_{R_d=1}$ is the distance from the crack tip where the $R_d$ ratio is equal to unity. As expected, a higher degree of plastic dissipation in the solid due to the combination of material properties ($l = 1$ μm and $N = 0.075$) and a mode II plane stress state, postpone the boundary of the equilibrium state of dislocation densities relative to a case where plasticity is reduced, such as $l = 20$ μm and $N = 0.4$ and mode I plane strain conditions. A preliminary analysis reveals that there is a strong relationship between the plastic SIF $K_P^{FEM}$ and the coordinate of the equilibrium state of dislocation densities $R_d$. This is further explored in Figs. 12c and 12d. As observed before (Fig. 5), a high value of the plastic SIF $K_P^{FEM}$ tends to be associated with a high GND density (and vice versa). The differences between plane strain and plane stress states with a variation in the hardening exponent $N$ are more significant for mode I, as shown in Figs. 12a and 12c. Figures 12b and 12d indicate that the stress-strain state (plane strain or plane stress) does not have a significant effect on the behaviour of the $R_d$ ratio under pure mode II.

One of the key novelties of the study by Jiang et al. [16] was to experimentally obtain the GND density maps. The authors mentioned that it is impossible to directly correlate the accumulated in-plane effective strain with the GND density distribution. Furthermore, the local progressive development of plastic strain is modulated by local strain hardening, typically due to back stresses associated with the accumulation of the measured GND density as well as the presence of SSDs, which are more difficult to measure. In the case of the crack tip with a finite radius of curvature as a result of the formed GND structure, the distribution of stress and strain may be more uniform relative to a sharp crack. Potentially, this is related to regions of high GND density ahead of the crack tips, indicating that the analysis of dislocation also plays a significant role in SGP theory. To this end, the introduced coordinate of the equilibrium state of the dislocation densities $R_d$ can serve as an additional basis for a conservative assessment of the outer boundary of the dominance region of gradient theories of plasticity.

**Conclusions**

We have comprehensively investigated the stationary crack tip fields of a material characterised by the mechanism-based strain gradient plasticity theory, which is grounded on Taylor's dislocation model. Our analysis spans both plane stress and plane strain stress states, as well as pure mode I and pure mode II loading conditions. Moreover, the interplay between plastic properties and strain gradient length scale is examined by mapping a wide range of values. A particular focus of this work is the study of the behaviour of dislocation

densities – angular and radial distributions of the density of geometrically necessary dislocations (GNDs) and statistically stored dislocations (SSDs) are computed, for varying stress states, fracture modes, hardening parameters and values of the intrinsic material length. The main findings of our work are as follows:

1. For a given remote elastic stress intensity factor, gradient effects appear to be stronger under plane stress conditions than under plane strain conditions. These differences are intrinsic to mode I fracture, with mode II conditions showing a very similar effective stress distribution ahead of the crack tip for both plane stress and plane strain states.

2. The numerical results reveal that the crack tip singularity depends on the fracture mode I/II. In all cases, the power of the stress singularity exceeds that of linear elasticity; under pure mode I values range from -0.543 to -0.672 while for mode II the range is -0.500 to -0.572. Also, crack tip fields do not have a separable solution and the crack tip stress fields exhibit a strong sensitivity to the work hardening exponent $N$ and the intrinsic material length scale $l$.

3. Nonlinear amplitude factor solutions for mechanism-based strain gradient plasticity are determined. These plastic stress intensity factors decrease with increasing material length scale $l$ and with decreasing strain hardening exponent $N$, being more sensitive to the latter.

4. The density of GNDs is large around the crack tip, but rapidly decreases away from the crack tip. On the contrary, the density of SSDs is not as large as that of GNDs around the crack tip, but decreases much more slowly away from it. The GND density magnitudes agree with experimental data except for distances very close to the crack tip, where measurements can be compromised by resolution issues.

5. A coupled effect of the fracture mode and the stress-strain state is identified on the dislocation density behaviour. In pure mode I, the GND density is located symmetrically with respect to the blunted crack tip. Under pure mode II, the GND density becomes concentrated in the blunted and sharpened parts of the crack tip. In this case, fracture is shown to be likely to initiate near the blunted region of the crack tip, where both the stress triaxiality and the GND density are at their maximum.

**Acknowledgements**

The authors ([1]VS, AT, RKh) gratefully acknowledge the financial support of the Russian Science Foundation under the Project 20-19-00158.